# Depth-Dependent EBIC Microscopy of Radial-Junction Si Micropillar Arrays

Kaden M. Powell[1] and Heayoung P. Yoon[1, 2]

[1] Electrical and Computer Engineering, University of Utah, Salt Lake City, UT 84112, USA

[2] Materials Science and Engineering, University of Utah, Salt Lake City, UT 84112, USA

Recent advances in fabrication have enabled radial-junction architectures for cost-effective and high-performance optoelectronic devices. Unlike a planar PN junction, a radial-junction geometry maximizes the optical interaction in the three-dimensional (3D) structures, while effectively extracting the generated carriers via the conformal PN junction. In this paper, we report characterizations of radial PN junctions that consist of *p*-type Si micropillars created by deep reactive-ion etching (DRIE) and an *n*-type layer formed by phosphorus gas diffusion. We use electron-beam induced current (EBIC) microscopy to access the 3D junction profile from the sidewall of the pillars. Our EBIC images reveal uniform PN junctions conformally constructed on the 3D pillar array. Based on Monte-Carlo simulations and EBIC modeling, we estimate local carrier separation/collection efficiency that reflects the quality of the PN junction. We find the EBIC efficiency of the pillar array increases with the incident electron beam energy, consistent with the EBIC behaviors observed in a high-quality planar PN junction. The magnitude of the EBIC efficiency of our pillar array is about 70 % at 10 kV, slightly lower than that of the planar device ($\approx$ 81 %). We suggest that this reduction could be attributed to the unpassivated pillar surface and the unintended recombination centers in the pillar cores introduced during the DRIE processes. Our results support that the depth-dependent EBIC approach is ideally suitable for evaluating PN junctions formed on micro/nanostructured semiconductors with various geometry.

# 1. INTRODUCTION

PN junctions are fundamental device elements that have been extensively used in various applications, including integrated electronic circuits, optical sensors and detectors, and energy harvesting and conversion systems (Chu et al., 2019; Sengupta et al., 1998; Neudeck, 1989). Recent advances in micro/nanofabrication have enabled three-dimensional (3D) architectures that offer design flexibility to produce high-performance optoelectronic devices using cost-effective semiconductors (Garnett and Yang, 2010; Li, 2012; Yoon et al., 2010; Um et al., 2015). These low-quality materials, however, exhibit short minority carrier diffusion lengths ($L_{n, p} < 10$ µm) due to high concentrations of impurities and structural defects (*e.g.*, point defect, vacancy, dislocation, grain boundary), limiting device performance designed in a planar geometry. In contrast, a radial-junction configuration maximizes light absorption along the length of the pillars while extracting minority carriers in the radial direction, effectively decoupling the competing processes. As a proof-of-concept, our previous work demonstrated over twofold higher power conversion efficiencies with radial junction solar cells compared to their planar junction counterparts (Yoon et al., 2010; Kendrick et al., 2010; Yoon et al., 2011).

Establishing robust junctions is essential for high-performance PN devices, as it controls the flow of excess carriers in one direction, but not the other, providing rectifying characteristics. Various fabrication methods have been proposed and demonstrated to construct the 3D structures with conformal junction formation using top-down (Dowling et al., 2017; Zeniou et al., 2014; Huang et al., 2011; Han et al., 2014; Qian et al., 2020) or bottom-up approaches (Lew and Redwing, 2003; Yoo et al., 2013). Plasma-based deep reactive ion etching (DRIE), also known as the "Bosch process" (Laermer and Schilp, 2003), has been widely used for 3D patterning owing to the fast and easy processing with reproducible structures. By repeating a cycle of plasma ion etching and conformal polymer coating, DRIE enables the production of high-aspect-ratio structures on various semiconductor substrates. However, the aggressive etching processes often cause unintended surface damage (*e.g.*, porous surface structures, electrically-active defect centers) (Oehrlein, 1989; B. Q. Wu et al., 2010). A shallow PN junction created on this (sub)surface exhibits poor rectification and inferior diode characteristics. Much of the research activity has been focused on fabrication approaches to reduce the damage and to enhance the fidelity of the 3D etched structures. The capability to access the local PN junctions and measure their characteristics is essential for further improving device performance.

Electron beam induced current (EBIC) microscopy is a powerful analytical technique for studying local electronic states of semiconductor materials and devices (Zhou et al., 2020; Leamy, 1982). EBIC uses a focused electron beam to create excess carriers (*i.e.*, electron-hole pairs) near a Schottky or a PN junction. The generated carriers in the quasi-neutral region are diffused in the ambipolar directions. The portion of carriers that reach the junction depends on the recombination rate of their travel path. Subsequently, the local built-in or applied electric field at the junction separates the electron-hole pairs, producing an induced current (*i.e.*, EBIC) in the external circuit. EBIC imaging is frequently used to map excess carrier recombination in semiconductors (Teplin et al., 2015). By fitting an EBIC profile that exponentially decays with distance from the junction, a minority carrier diffusion length can be determined (Yakimov, 2015). Moreover, recent studies have proposed advanced EBIC simulations and numerical modeling for quantitative analysis of convoluted EBIC signals (Zhou et al., 2020; Haney et al., 2016).

In this work, we report measurements of radial PN junctions formed on Si micropillar arrays based on depth-dependent EBIC microscopy. By controlling the beam energy that determines the interaction volume between the injected electron beam and the Si structures, we measure the EBIC characteristics of the pillar surface as well as the pillar cores. Our radial PN junctions consist of *p*-type Si micropillar cores fabricated by DRIE and an *n*-type Si shell formed by phosphorous diffusion. We perform EBIC from the sidewall of the pillars to visualize the conformal PN junction. The junction quality of the micropillar array is determined by extracting a local carrier separation/collection efficiency using simulations and EBIC modeling. We compare the EBIC efficiency of the radial junction to the baseline planar PN junction, providing qualitative and quantitive assessment.

## 2. MATERIALS AND METHODS

*a. Fabrication of Si Micropillar Arrays*

The radial PN junctions studied in this work consist of *p*-type Si micropillar cores fabricated by DRIE and an *n*-type Si shell formed by phosphorous gas diffusion. Details of full fabrication steps can be found elsewhere (Yoon et al., 2010; Yoon et al., 2011). Briefly, the process was begun by conventional lithography to pattern a 200 nm thermal oxide layer on Si (*p*-type; $\rho$ = 0.01 ~ 0.02 $\Omega$·cm). This patterned $SiO_2$ layer serves as an etch mask in the subsequent anisotropic Si etching. In DRIE, we used the repeated cycles of sulfur hexafluoride ($SF_6$) etching and octafluorocyclobutane ($C_4F_8$)/oxygen ($O_2$) polymer coating (Figure 1a). Each cycle is composed of (i) 3.5 s $SF_6$, (ii) 1.5 s $C_4F_8$ + $O_2$, and (iii) 1 s $O_2$ at an RF power of 1500 W, providing a Si etch rate of about 2.5 μm/min. To remove the sidewall roughness of the pillar arrays, we used several steps of cleaning processes: (i) $O_2$ plasma cleaning (200 sccm $O_2$ flow rate, 30 mTorr chamber pressure, and 1800 W RF power for 5 min), (ii) Piranha cleaning ($H_2SO_4$ : $H_2O_2$ = 1 : 1), and (iii) two successive wet oxidation (1000 °C for 25 min, resulting in ≈ 300 nm thick $SiO_2$) and oxide removal processes (10 % HF for 1 min). To form radial PN junctions, we used gas phase diffusion of a phosphorus oxychloride ($POCl_3$) source at 1000 °C for 13 minutes. A 300 nm-thick aluminum (Al) metal film was thermally evaporated on the backside of the *p*-type Si and annealed at 600 °C for 10 minutes in the $N_2$ atmosphere. The frontside contacts to the $n^+$ diffused layers were formed with indium dots on the four corners of the pillar arrays after a native oxide removal in buffered oxide etchant (BOE 1:50, 30 s).

*b. Si planar PN junction controls*

A high-quality Si planar PN device served as a baseline control to evaluate our radial junctions formed on the DRIE-fabricated Si micropillar arrays. We purchased a batch of commercially available Si solar cells (Solar Made) fabricated with high-quality, single-crystalline Si materials. The surface of each device was passivated (*e.g.*, SiN) to reduce the surface recombination velocity. Previously, we conducted a cross-sectional EBIC measurement using this sample. The obtained carrier separation/collection efficiency near the planar PN junction was close to 100 %, suggesting a robust PN junction of the planar device (Yoon et al., 2014). In this work, we perform EBIC in a depth-dependent configuration.

*c. EBIC Characterizations*

EBIC measurements were carried out in a scanning electron microscope (SEM) equipped with a nano-manipulator. This probe arm was used for the placement of an electrical probe on a metal contact of the device, whereas the bottom metal contact was earthed to the SEM stub. The electrical wires were fitted with coaxial electrical feedthrough of the SEM, connecting to an EBIC current amplifier. Once the electrical contacts were made, a series of EBIC/SEM images were acquired at different incident beam energies using a software package.

*d. Monte Carlo Simulations (e-beam)*

We used numerical simulations to estimate the two-dimensional (2D) profile in Si at various incident beam energies (CASINO software package: Monte CArlo SImulation of electroN trajectory in sOlids) (Drouin et al., 2007). In each simulation, we used 200,000 electrons at a fixed accelerating voltage in the range of 1 kV to 30 kV. The injected electrons were propagated in the Si matrix via elastic and inelastic scatterings until their energy becomes 50 eV or less. The radius of the normal incident beam was set to 2 nm. A homogeneous Si substrate in a planar geometry was used for the simulation.

**3. RESULTS AND DISCUSSION**

Figure 1 displays the key fabrication steps to create Si micropillar arrays. Deep reactive ion etching (DRIE) is a well-established fabrication technique to produce 3D structures by repeating the cycle of a plasma ion etching and a conformal polymer coating (Figure 1a). These etching/coating processes, however, introduce unintended porous surface structures and electrically-active defect centers. An example of the scalloping profile of the etched structures is shown in Figure 1b. To remove this structural damage, we used two successive thermal oxidation (1,000 °C for 25 min; ≈ 300 nm thick $SiO_2$; Figure 1c) and oxide removal processes (10 % HF for 1 min). A representative SEM image in Figure 1d confirms the dramatically enhanced surface smoothness of the Si micropillars after the rigorous cleaning and oxidation/strip processes. We formed the radial PN junction using gas diffusion of *n*-type phosphorous dopants. An estimated surface doping concentration ($N_d$) is $\approx 10^{20}$ cm$^{-3}$ with a junction depth ($x_j$) of $\approx 0.3$ μm (Neudeck, 1989). The complete Si micropillars measured $\approx 30$ μm in height and $\approx 7$ μm in diameter with a distance between the pillars of approximately 4 μm (Figure 1e). The current-voltage (*I-V*) curves of the pillar array in Figure 1f showed a good diode behavior. We estimated the turn-on voltage of 0.63 V, the leakage current of 10 nA, and the ideality factor of about 1.7. Comprehensive dark and light *I-V* characteristics of the radial junctions in various geometrical parameters (*e.g.*, diameter, height, pillar-to-pillar-distance) can be found elsewhere (Yoon et al., 2010; Yoon et al., 2011). While extremely informative, *I-V* curves reflect the overall PN junction performance and do not capture the local junction properties.

EBIC microscopy allows a direct access to the local PN junction in 3D with an adjustable probe size from 10's nm to several μm, in accordance with the interaction volume between the incident electron beam and the semiconductors. To visualize the local junctions at the level of individual pillars, we carefully cleaved the Si pillar array using a fine scriber and exposed the cross-section. The sample was mounted on an EBIC holder, where the metal contacts of the emitter (*i.e.*, indium dots on $n^+$-Si) and the collector (*i.e.*, Al on *p*-Si) were connected to the external EBIC circuit. The electron beam was injected from the *n*-Si emitter shell to the PN depletion region and the *p*-Si pillar core. Figure 2 displays an SEM and the corresponding 5keV (Figure 2b) and 10 keV (Figure 2c) EBIC images. Figure 2d shows the overlaid SEM and 5 keV EBIC, indicating a continuous PN junction formed across the 3D geometry. The overall EBIC intensities of the individual pillars at 5 keV and 10 keV are relatively uniform, suggesting the presence of conformal radial junctions along the individual micropillars.

For a quantitative analysis, we extract the EBIC line scans along/across the pillars. The line scan plot across the pillar diameter (Figure 2e) shows the highest EBIC value near the pillar center ($\approx$ 80 nA) that decreases gradually with the electron beam probe moving away to the perimeter of the pillar ($\approx$ 50 nA). Considering the direction of the incident electron beam to the curved pillar surface, as illustrated in the inset of Figure 2e, we suggest that this EBIC change is mainly attributed to the shape of the pillar rather than due to inhomogeneous junction properties. With the increase of the backscattered electrons (BSE) and the decrease of the effective electron-hole pair (EHP) generation volume at the curved pillar surface, the reduction of the EBIC magnitude is evident near the pillar perimeter.

Figure 2f displays the EBIC line scans along the length of the pillars (*i.e.*, axial-direction), showing a relatively constant EBIC value within the pillars at a fixed accelerating voltage. The mean EBIC value of the pillar increases from 80 nA at 5 keV to 480 nA at 10 keV. We observe the highest EBIC values are present in the area of the pillar base, which is associated with the direct local carrier generation within the depletion region. When the electron beam is directly injected to the cross-sectional PN junction (*i.e.*, mechanically cleaved junction area), the generated EHPs are separated quickly without diffusion owing to the built-in electric field (C. J. Wu and Wittry, 1978; Yakimov, 2015). In contrast, the electron beam irradiated on the pillar sidewalls generates the EHPs in the depletion region as well as the charge-neutral regions (*i.e.*, *n*-Si shell, *p*-Si pillar core). The excess carriers must travel to the junction (*i.e.*, ambipolar diffusion) before they are separated and collected in EBIC. Since the *n*-Si emitter layer ($N_d \approx$ $10^{20}$ cm$^{-3}$, $x_j \approx$ 300 nm) of our pillars is conductive, yet highly defective, the EHPs generated in this region tend to be recombined, decreasing overall EBIC values as compared to the direct electron beam injection at the cross-sectional PN junction. As the EHP generation volume increases with the accelerating voltage (*i.e.,* 5 keV to 10 keV), the portion of the EHPs generated in the *n*-Si decreases, resulting in a comparable EBIC value of the pillar and near the base.

To assess the local junction quality of the micropillar array, we collected the baseline EBIC characteristics of a commercial planar device (Solar Made). This planar PN junction ($n^+$-*p*) was built on a high-purity single crystalline Si substrate, and it showed a carrier collection efficiency close to 100 % in the depletion region obtained in a normal collector EBIC configuration (Yoon et al., 2014). Figure 3 displays a representative SEM image of the planar PN device and the corresponding EBIC maps collected at 5 keV, 10 keV, and 20 keV. The large dark area of the EBIC images is associated with the metal contact, highlighted in yellow in the SEM image (Figure 3a). The injected electron beam (1 keV ~ 30 keV) does not penetrate this thick metal layer (a few mm thick Ag paste), producing negligible EBIC signals (Figure 3b ~ 3d). The dark speckles in the 5 keV EBIC image are likely attributed to thin organic residue or dust particles on the sample, of which EBIC contribution becomes insignificant with the higher beam energies (> 10 keV). Qualitatively, the bright contrast increases with a higher keV, showing similar behaviors as those observed with the pillar array radial junction (Figure 2).

Figures 3(f) through 3(g) show the representative line scans extracted from the EBIC images (Figures 3b ~ 3d). A relatively constant EBIC was observed in the device area, a stack of *p*-Si collector, *n*-Si emitter, and SiN passivation layer. A notable current fluctuation near the metal contact is mainly attributed to the spread of the metal paste. By aligning the line scans, we find that a low keV EBIC is much more sensitive to the surface features than higher keV. For instance, two distinct peaks observed in the 5 keV EBIC line scan (marked with a green box) conform to the sample topography shown in SEM (Figure 3a). This feature becomes less distinguishable with increasing incident beam energy, as the electron beam penetrates deep in the sample with a larger EHP generation volume. We used the EBIC images from 5 keV to 30 keV and calculated mean EBIC values for the Si area, shown in Figure 3e. The increase of EBIC with a higher keV is evident in the line scans. The average EBIC value increases from 127 nA at 5kV to 3.55 µA at 30 kV, increasing over one order of magnitude. Interestingly, the EBIC values observed in the planar junction are slightly higher than those in the radial junction in Figure 2: 127 nA (*vs*. 83 nA of the radial junction) at 5 keV, 574 nA (*vs*. 492 nA of the radial junction) at 10 keV.

The experimental results qualitatively suggest that EBIC magnitude near the PN junctions is strongly influenced by the EHP generation by the incident electron beam and the local carrier separation/collection properties. To gain a deeper understanding of the local radial junction characteristics, we estimate the carrier generation profile using Monte Carlo simulations and calculate the local carrier collection efficiency for the planar and the radial junctions. Figure 4a (top) displays an example of the simulated electron trajectories, where a ray of 5 keV electron beam is irradiated onto a Si substrate. The blue lines represent the collision events of the primary electrons with Si until they lose their initial energy (*i.e.*, 5,000 V in this example) to 50 V or lower. The red lines represent the paths of the backscattered electrons. A corresponding energy contour plot is shown in the bottom image. For instance, the 95 % contour represents the sample area where the injected primary electrons have lost 95 % of their initial energy. Figure 4b plots the estimated interaction bulb size at different accelerating voltages (1 kV ~ 30 kV). The overall ratio of depth to diameter (depth/diameter) is comparable for higher lost-energy contours (> 75 %), yet slightly higher for low energy contours (< 50 %), indicating a pear-shape of the interaction bulb. The calculated bulb size at 1 keV is approximately 19 nm, inferring that the spatial resolution of the EBIC image for flat Si devices can be achieved as high as < 20 nm. The inset of Figure 4b shows the increase of the penetration depth with the beam energy, which was extracted from the 95 % energy contour of each simulation. The numerical fit overlaid on the datasets confirms the bulb size is proportional to $E_b^{1.78}$ ($E_b$ is the beam energy), showing an excellent agreement with the analytical prediction of $\approx E_b^{1.7}$ by Wittry *et al.* (Wittry and Kyser, 1967). By controlling the incident beam energy, the size of the EHP generation bulb can be tunable from 10's nm to several μm, offering versatility to study local carrier dynamics in optoelectronic semiconductor materials and devices.

Based on the Monte-Carlo simulations and EBIC modeling (Leamy, 1982; Haney et al., 2016; Yakimov, 2015), we estimated the local carrier separation/carrier efficiency of the radial junction and compared it to planar PN controls. The EBIC collection efficiency ($\eta_{EBIC}$) is defined as the ratio of the measured current ($I_{EBIC}$) to the EHP generation rate (β). Here, *e* is the unit charge (1.6×10$^{-19}$ C).

$$\eta(EBIC) = \frac{I_{EBIC}}{e \cdot \beta} \quad (1)$$

The generation rate, which is the total number of EHPs created by the injected electron beam, can be calculated as below.

$$\beta = \frac{I_b \cdot E_b \cdot \alpha}{e \cdot E_{EHP}} \qquad (2)$$

$E_b$ is the incident electron beam energy, $\alpha$ is the fraction of beam energy absorbed inside the material (*i.e.*, Si in our case), and $E_{EHP}$ is the average energy to create an electron-hole pair. The $I_b$ of our SEM was measured in the range of 250 pA ($E_b$ = 5 kV) to 300 pA ($E_b$ = 20 kV). We calculated the magnitude of $\alpha$ using the backscattered coefficient obtained from the Monte Carlo simulation (*e.g.*, 0.152 at 5 keV, 0.142 at 20 keV). The $E_{EHP}$ was estimated using an empirical relation of $E_{EHP} = 2.596 E_g + 0.174$ (Kobayashi et al., 1972), giving $E_{EHP} \approx 3.621 eV$ for Si ($E_g = 1.12 eV$). The EBIC currents ($I_b$) extracted from the line scans in Figures 2 and 3 were used for the pillar array and the planar device, respectively. We note that a typical uncertainty in our EBIC measurement and analysis is about 10 % associated with the fluctuations of the baseline e-beam current ($I_b$) and the signal-to-noise ratio of the EBIC preamplifier. Also, the parameters extracted from the Monte Carlo simulations (*e.g.*, backscattered coefficient, mean EHP generation rate (*β*), empirical parameter ($E_{EHP}$) to generate EHPs) contribute to the uncertainty.

Finally, we plot the resulting EBIC efficiency of the devices at different incident beam energies in Figure 4c. The EBIC efficiency increases with the incident beam energy, reaching close to unity at $E_b$ > 15 keV for the planar PN junction device. A similar trend was observed for the radial junction of the pillar array, yet the overall EBIC efficiency is slightly lower than that of the planar device (about 10 %). In both cases, EBIC was measured in the depth-dependent configuration. The injected electrons travel from the highly-doped emitter (a few 100 nm thick) to the depletion region (< 1 μm) and the *p*-Si collector, generating the EHPs in three different layers. The low EBIC efficiency at 5 keV (50 % for the planar device; 30 % for the pillar array) is likely attributed to the EHP production in the highly-doped emitter region. Our Monte Carlo simulation shows an interaction bulb size of $(300 \text{ nm})^3$ at 5 keV, suggesting that most EHPs were produced in the highly-defective (*i.e.*, high-density of recombination centers) emitter region that promotes excess carrier recombination. At higher keV, most EHPs are generated in the strong built-in electric field region and the collector, increasing the EBIC efficiency. The magnitude of the EBIC efficiency of our pillar array is about 70 % at 10 kV, slightly lower than that of the planar device (≈ 81 %). We speculate that a slightly higher EBIC efficiency for the planar junction is associated with the surface passivation (*e.g.*, SiN) that decreases the surface recombination of EHPs. Alternatively, the material quality of the *p*-Si pillar core (≈ $10^{18}$ cm$^{-3}$) is less defective than their *n*-Si shell (≈ $10^{20}$ cm$^{-3}$). Nevertheless, our observation indicates that the surface damage introduced on the pillars by DRIE could be effectively removed by rigorous cleaning and oxidation/strip processes.

## 4. SUMMARY AND CONCLUSIONS

In summary, we have examined the radial junction characteristics of Si micropillar arrays using depth-dependent EBIC microscopy. The EBIC images collected from the sidewall of the pillars confirm the uniform PN junction conformally constructed on the 3D pillar array. We find the EBIC efficiency of the pillar array increases with the injected electron-beam voltage, consistent with the EBIC behaviors observed in a high-quality planar PN junction. The magnitude of the EBIC efficiency of our pillar array is about 70 % at 10 kV, slightly lower than that of the planar device (≈ 81 %). We suggest that this reduction could be attributed to the unpassivated pillar surface or the low material quality of the pillar core. Our results support that the depth-dependent EBIC approach is ideally suitable for evaluating 3D conformal PN junctions formed on micro/nanostructures with various geometries.

KEY WORDS

EBIC, Electron Beam Induced Current, EBIC efficiency, Radial Junction, PN Junction, Micropillar Array, Solar Cells, Local Characterization, Scanning Electron Microscopy

ABBREVIATIONS

2D: two-dimensional; 3D: three-dimensional; BSE: Backscattered electron; CASINO: Monte-CArlo SImulation of electroN trajectory in sOlids; DRIE: Deep reactive-ion etching; EBIC: Electron-beam induced current; EHP: Electron-hole pair; RF: Radio frequency; SEM: Scanning electron microscopy

ACKNOWLEDGEMENTS

The authors acknowledge the support from B. Baker, D. Magginetti, and S. Pritchett for the development of device fabrication. The radial junction processes and the EBIC data acquisition were performed at Penn State University (University Park, PA, USA) and the National Institute of Standards and Technology (Gaithersburg, MD, USA). We thank Y. Yuwen, T. Mayer, P. Haney, and N. Zhitenev for valuable discussions.

FUNDING

This research was supported by a University of Utah Seed Grant and New Faculty Start-up Funds. We acknowledge support by the USTAR shared facilities at the University of Utah, in part, by the MRSEC Program of NSF under Award No. DMR-1121252.

AVAILABILITY OF DATA AND MATERIALS

The datasets used in this study are available from the corresponding author on reasonable request.

AUTHORS' CONTRIBUTIONS

KMP and HPY have contributed to sample preparation, data acquisition, data analysis, and manuscript writing. HPY supervised the overall project. All authors read and approved the final manuscript.

COMPETING INTEREST

The authors declare that they have no competing interests.

**Figure 1.**

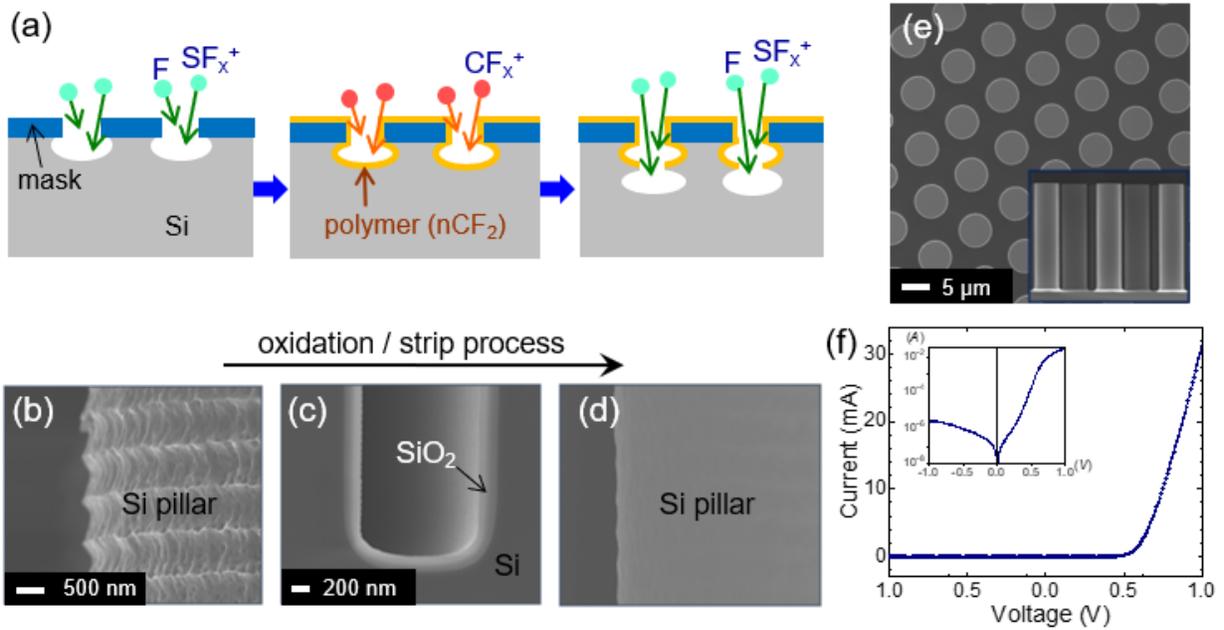

(a) Schematics illustrating a single cycle of the deep reactive-ion etching (DRIE) process. (b-d) SEM images of a pillar sidewall after (b) DRIE, (c) thermal oxidation, and (d) HF oxide removal. (e) Top-view SEM image of a portion of the complete Si micropillar array. Inset displays cross-section of the device. (f) Typical *I-V* curve of pillar array radial junctions, showing ratifying diode behaviors. Inset plots the *I-V* on a log scale.

**Figure 2.**

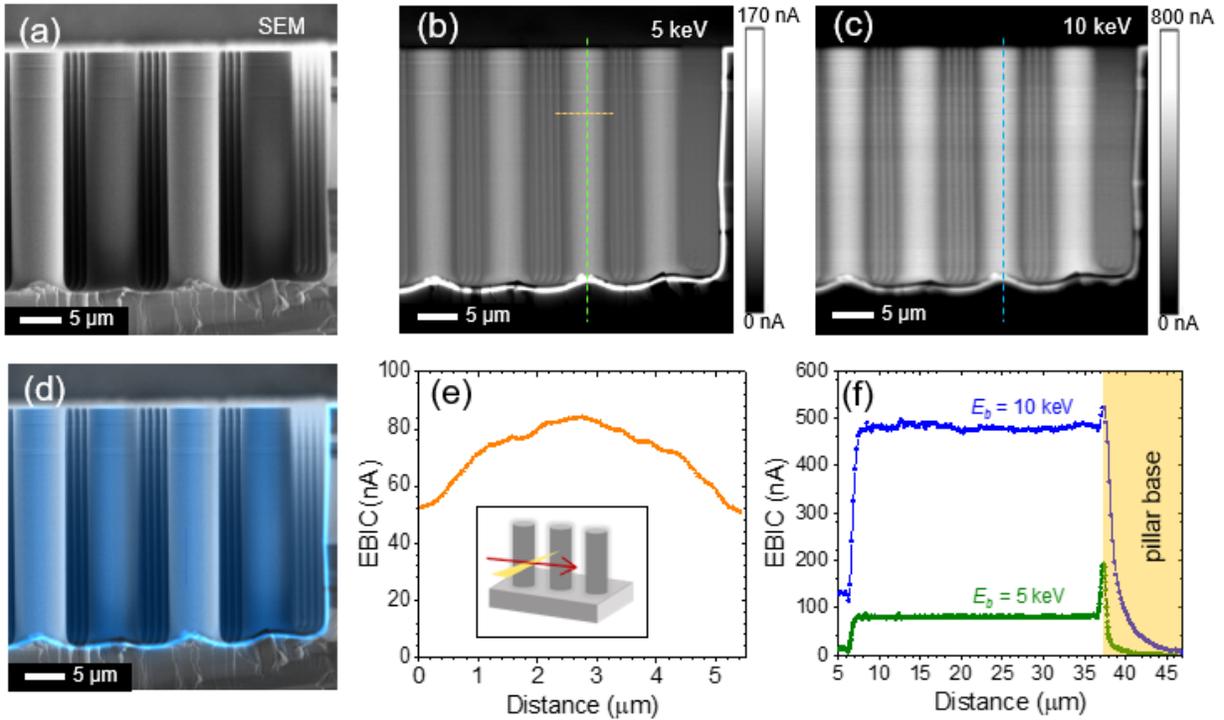

(a) SEM image of micropillar array and corresponding EBIC (electron beam induced current) maps at (b) 5 keV and (c) 10 keV. The relative signal uniformity suggests the presence of conformal radial junctions of individual pillars. (d) Overlay of SEM with 5 keV EBIC image confirms continuous PN junction across device. (e) Extracted EBIC line scan across pillar diameter. (f) EBIC line scans at 5 keV and 10 keV along pillar lengths demonstrate uniform signal magnitude and higher EHP generation at higher energy. Peak EBIC values occur in the pillar base region, where the cross-sectional PN junction is exposed.

**Figure 3.**

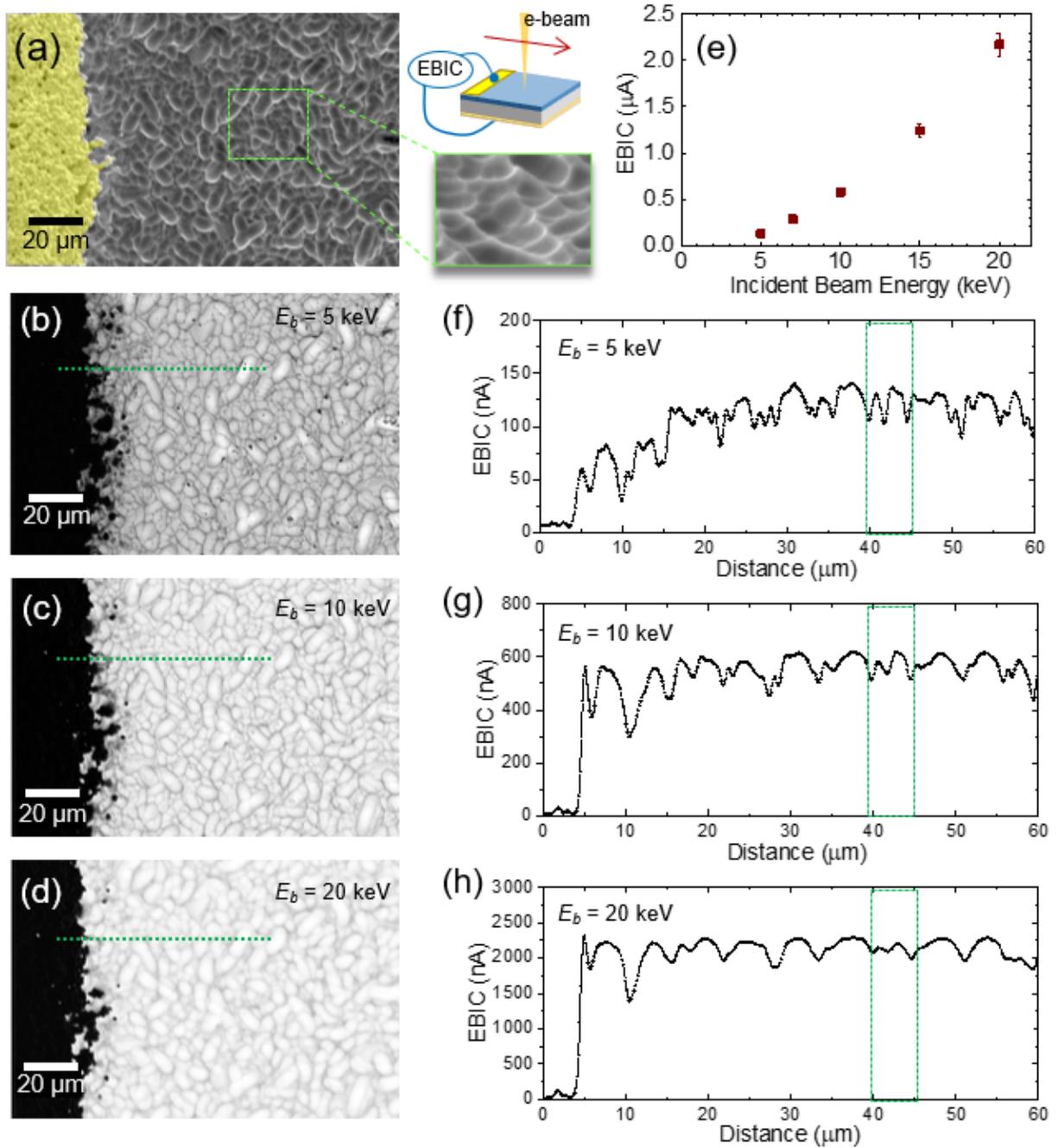

(a) SEM image of planar PN device with metal contact highlighted in yellow. [center top] Schematic of EBIC measurement. [center bottom] Slightly tilted SEM image of the sample, illustrating surface roughness. Corresponding EBIC scans of the planar device at (b) 5 keV, (c) 10 keV, and (d) 20 keV showing higher contrast and reduced spatial resolution with higher energy. (e) Increasing mean EBIC current calculated for Si line scans. (f-h) Plots of EBIC current along line scans shown with green lines in (b-d): (f) 5 keV, (g) 10 keV, (h) 20 keV. The green box highlights the representative sample topography. The two distinct features at 5 keV become broad and indistinguishable with higher beam energies.

**Figure 4.**

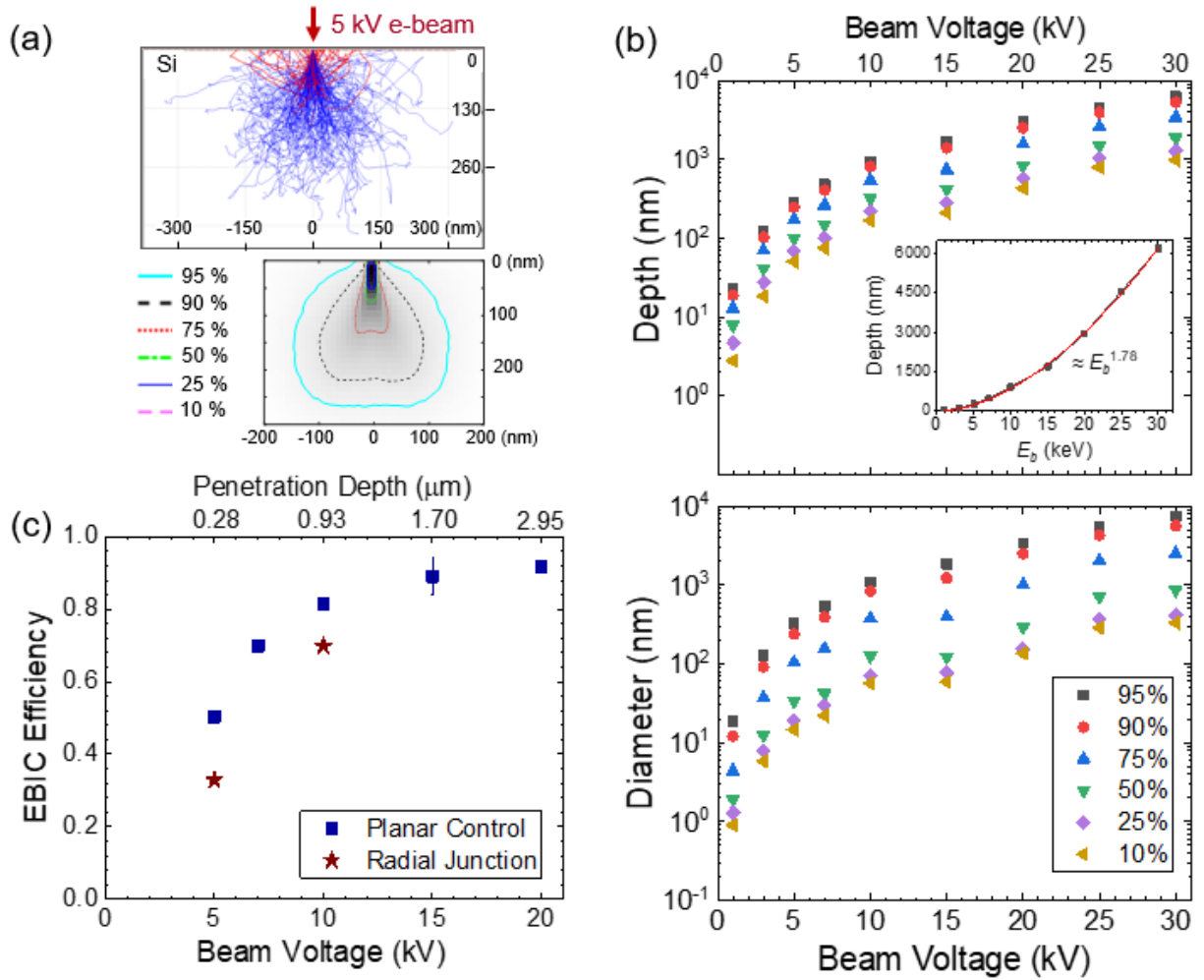

(a) Simulated electron paths for a ray of 5 kV electron beam in Si (200,000 electrons). Blue lines represent collision events of electrons with Si. Red lines represent backscattered electrons. (a, bottom) Energy contour plot from 5 kV simulation with percent energy loss contours. (b) Simulated interaction bulb size at various accelerating voltages: [top] maximum depth and [bottom] maximum diameter. Inset shows the penetration depth with beam energy (95 % energy contour) and corresponding curve fit. (c) Estimated EBIC efficiency of the radial junction compared to the planar PN junction control.

# REFERENCES

Chu, Y. H., Qian, C. Q., Chahal, P., & Cao, C. Y. (2019). Printed Diodes: Materials Processing, Fabrication, and Applications. *Advanced Science*, 6(6). doi:ARTN 1801653 10.1002/advs.201801653.

Dowling, K. M., Ransom, E. H., & Senesky, D. G. (2017). Profile Evolution of High Aspect Ratio Silicon Carbide Trenches by Inductive Coupled Plasma Etching. *Journal of Microelectromechanical Systems*, 26(1), 135-142. doi:10.1109/Jmems.2016.2621131.

Drouin, D., Couture, A. R., Joly, D., Tastet, X., Aimez, V., & Gauvin, R. (2007). CASINO V2.42 - A fast and easy-to-use modeling tool for scanning electron microscopy and microanalysis users. *Scanning*, 29(3), 92-101. doi:DOI 10.1002/sca.20000.

Garnett, E., & Yang, P. D. (2010). Light Trapping in Silicon Nanowire Solar Cells. *Nano Letters*, 10(3), 1082-1087. doi:10.1021/nl100161z.

Han, H., Huang, Z. P., & Lee, W. (2014). Metal-assisted chemical etching of silicon and nanotechnology applications. *Nano Today*, 9(3), 271-304. doi:10.1016/j.nantod.2014.04.013.

Haney, P. M., Yoon, H. P., Gaury, B., & Zhitenev, N. B. (2016). Depletion region surface effects in electron beam induced current measurements. *Journal of Applied Physics*, 120(9). doi:Artn 095702 10.1063/1.4962016.

Huang, Z. P., Geyer, N., Werner, P., de Boor, J., & Gosele, U. (2011). Metal-Assisted Chemical Etching of Silicon: A Review. *Advanced Materials*, 23(2), 285-308. doi:10.1002/adma.201001784.

Kendrick, C. E., Yoon, H. P., Yuwen, Y. A., Barber, G. D., Shen, H. T., Mallouk, T. E., et al. (2010). Radial junction silicon wire array solar cells fabricated by gold-catalyzed vapor-liquid-solid growth. *Applied Physics Letters*, 97(14). doi:Artn 143108 10.1063/1.3496044.

Kobayashi, T., Koyama, M., Sugita, T., & Takayanagi, S. (1972). Performance of Gaas Surface-Barrier Detectors Made from High-Purity Gallium-Arsenide. *Ieee Transactions on Nuclear Science*, Ns19(3), 324-+. doi:Doi 10.1109/Tns.1972.4326745.

Laermer, F., & Schilp, A. (2003). Method of anisotropic etching of silicon. *Patent US6531068 (US)*.

Leamy, H. J. (1982). Charge Collection Scanning Electron-Microscopy. *Journal of Applied Physics*, 53(6), R51-R80. doi:Doi 10.1063/1.331667.

Lew, K. K., & Redwing, J. M. (2003). Growth characteristics of silicon nanowires synthesized by vapor-liquid-solid growth in nanoporous alumina templates. *Journal of Crystal Growth*, 254(1-2), 14-22. doi:10.1016/S0022-0248(03)01146-1.

Li, X. L. (2012). Metal assisted chemical etching for high aspect ratio nanostructures: A review of characteristics and applications in photovoltaics. *Current Opinion in Solid State & Materials Science*, 16(2), 71-81. doi:10.1016/j.cossms.2011.11.002.

Neudeck, G. W. (1989). The PN junction diode. Reading, Mass.: Addison-Wesley.

Oehrlein, G. S. (1989). Dry Etching Damage of Silicon - a Review. *Materials Science and Engineering B-Solid State Materials for Advanced Technology*, 4(1-4), 441-450. doi:Doi 10.1016/0921-5107(89)90284-5.

Qian, Y., Magginetti, D. J., Jeon, S., Yoon, Y., Olsen, T. L., Wang, M., et al. (2020). Heterogeneous Optoelectronic Characteristics of Si Micropillar Arrays Fabricated by Metal-Assisted Chemical Etching. https://ui.adsabs.harvard.edu/abs/2020arXiv200616308Q. Accessed June 01, 2020.

Sengupta, D. L., Sarkar, T. K., & Sen, D. (1998). Centennial of the semiconductor diode detector. *Proceedings of the Ieee*, 86(1), 235-243. doi:Doi 10.1109/5.658775.

Teplin, C. W., Grover, S., Chitu, A., Limanov, A., Chahal, M., Im, J., et al. (2015). Comparison of thin epitaxial film silicon photovoltaics fabricated on monocrystalline and polycrystalline seed layers on glass. *Progress in Photovoltaics*, 23(7), 909-917. doi:10.1002/pip.2505.

Um, H. D., Kim, N., Lee, K., Hwang, I., Seo, J. H., Yu, Y. J., et al. (2015). Versatile control of metal-assisted chemical etching for vertical silicon microwire arrays and their photovoltaic applications. *Scientific Reports*, 5. doi:ARTN 11277 10.1038/srep11277.

Wittry, D. B., & Kyser, D. F. (1967). Measurement of Diffusion Lengths in Direct-Gap Semiconductors by Electron-Beam Excitation. *Journal of Applied Physics*, 38(1), 375-&. doi:Doi 10.1063/1.1708984.

Wu, B. Q., Kumar, A., & Pamarthy, S. (2010). High aspect ratio silicon etch: A review. *Journal of Applied Physics*, 108(5). doi:Artn 051101 10.1063/1.3474652.

Wu, C. J., & Wittry, D. B. (1978). Investigation of Minority-Carrier Diffusion Lengths by Electron-Bombardment of Schottky Barriers. *Journal of Applied Physics*, 49(5), 2827-2836. doi:Doi 10.1063/1.325163.

Yakimov, E. B. (2015). What is the real value of diffusion length in GaN? *Journal of Alloys and Compounds*, 627, 344-351. doi:10.1016/j.jallcom.2014.11.229.

Yoo, J., Dayeh, S. A., Tang, W., & Picraux, S. T. (2013). Epitaxial growth of radial Si p-i-n junctions for photovoltaic applications. *Applied Physics Letters*, 102(9). doi:Artn 093113 10.1063/1.4794541.

Yoon, H. P., Haney, P. M., Schumacher, J., Siebein, K., Yoon, Y., & Zhitenev, N. B. (2014). Effects of Focused-Ion-Beam Processing on Local Electrical Measurements of Inorganic Solar Cells. *Microscopy and Microanalysis*, 20(S3), 544-545. doi:10.1017/S1431927614004449.

Yoon, H. P., Yuwen, Y. A., Kendrick, C. E., Barber, G. D., Podraza, N. J., Redwing, J. M., et al. (2010). Enhanced conversion efficiencies for pillar array solar cells fabricated from crystalline silicon with short minority carrier diffusion lengths. *Applied Physics Letters*, 96(21). doi:Artn 213503 10.1063/1.3432449.

Yoon, H. P., Yuwen, Y. A., Shen, H., Podraza, N. J., Mallouk, T. E., Dickey, E. C., et al. (2011). Parametric study of micropillar array solar cells. *37th IEEE Photovoltaic Specialists Conference*, 000303-000306. doi:10.1109/PVSC.2011.6185905.

Zeniou, A., Ellinas, K., Olziersky, A., & Gogolides, E. (2014). Ultra-high aspect ratio Si nanowires fabricated with plasma etching: plasma processing, mechanical stability analysis against adhesion and capillary forces and oleophobicity. *Nanotechnology*, 25(3). doi:Artn 035302 10.1088/0957-4484/25/3/035302.

Zhou, R. N., Yu, M. Z., Tweddle, D., Hamer, P., Chen, D., Hallam, B., et al. (2020). Understanding and optimizing EBIC pn-junction characterization from modeling insights. *Journal of Applied Physics*, 127(2). doi:Artn 024502 10.1063/1.5139894.